\documentclass[%
 reprint,
superscriptaddress,
 amsmath,amssymb,
 aps,
floatfix,
]{revtex4-2}

\usepackage{graphicx}
\usepackage{dcolumn}
\usepackage{bm}
\usepackage{braket}
\usepackage{siunitx}
\usepackage{bbold}
\DeclareSIUnit{\belmilliwatt}{Bm}
\DeclareSIUnit{\dBm}{\deci\belmilliwatt}
\usepackage[T1]{fontenc}
\usepackage{placeins}
\usepackage[utf8]{inputenc}

\usepackage{hyperref}
\hypersetup{
    colorlinks=true,
    linkcolor=magenta,
    citecolor=magenta,
    filecolor=magenta,      
    urlcolor=magenta,
}

\begin{document}

\title{Hotter is easier: unexpected temperature dependence of spin qubit frequencies}

\author{Brennan~Undseth}
 \thanks{These authors contributed equally to this work.}
\author{Oriol~Pietx-Casas}
 \thanks{These authors contributed equally to this work.}
\author{Eline~Raymenants}
\author{Mohammad~Mehmandoost}
\author{Mateusz~T.~M\k{a}dzik}
\author{Stephan~G.J.~Philips}
\author{Sander~L.~de~Snoo}
\affiliation{QuTech and Kavli Institute of Nanoscience, Delft University of Technology, Lorentzweg 1, 2628 CJ Delft, The Netherlands}
\author{David~J.~Michalak}
\author{Sergey~V.~Amitonov}
\author{Larysa~Tryputen}
\affiliation{QuTech and Netherlands Organization for Applied Scientific Research (TNO), Stieltjesweg 1, 2628 CK Delft, Netherlands}
\author{Brian~Paquelet~Wuetz}
\author{Viviana~Fezzi}
\author{Davide~Degli~Esposti}
\affiliation{QuTech and Kavli Institute of Nanoscience, Delft University of Technology, Lorentzweg 1, 2628 CJ Delft, The Netherlands}
\author{Amir~Sammak}
\affiliation{QuTech and Netherlands Organization for Applied Scientific Research (TNO), Stieltjesweg 1, 2628 CK Delft, Netherlands}
\author{Giordano~Scappucci}
\author{Lieven~M.~K.~Vandersypen}
\affiliation{QuTech and Kavli Institute of Nanoscience, Delft University of Technology, Lorentzweg 1, 2628 CJ Delft, The Netherlands}

\begin{abstract}

As spin-based quantum processors grow in size and complexity, maintaining high fidelities and minimizing crosstalk will be essential for the successful implementation of quantum algorithms and error-correction protocols. In particular, recent experiments have highlighted pernicious transient qubit frequency shifts associated with microwave qubit driving. Workarounds for small devices, including prepulsing with an off-resonant microwave burst to bring a device to a steady-state, wait times prior to measurement, and qubit-specific calibrations all bode ill for device scalability. Here, we make substantial progress in understanding and overcoming this effect. We report a surprising non-monotonic relation between mixing chamber temperature and spin Larmor frequency which is consistent with observed frequency shifts induced by microwave and baseband control signals. We find that purposefully operating the device at \SI{200}{\milli\kelvin} greatly suppresses the adverse heating effect while not compromising qubit coherence or single-qubit fidelity benchmarks. Furthermore, systematic non-Markovian crosstalk is greatly reduced. Our results provide a straightforward means of improving the quality of multi-spin control while simplifying calibration procedures for future spin-based quantum processors.

\end{abstract}

\maketitle 

\section{Introduction}

A major benefit of using semiconductor spins as the building blocks of a large-scale quantum computer is that they do not strictly require dilution refrigerator base temperatures (\SI{10}{}-\SI{20}{\milli\kelvin}). Demonstrations of qubit operation above \SI{1}{\kelvin} for electron spins confined in Si-MOS \cite{yang2019hotsilicon, Petit_2020hot}, and recently at \SI{4}{\kelvin} for a hole spin confined in a FinFET \cite{Camenzind_2022hot}, are encouraging steps towards the realization of classical and quantum hardware coexisting at the same temperature.

However, several factors motivate the operation of spin qubits at lower temperatures of about \SI{10}{\milli\kelvin}. These include the desires to limit charge noise as well as minimize electron temperature for widely used readout methods \cite{Petit_2018_chargenoise,Mills_2022_Elzerman}. The intuitive benefits of minimizing the environmental temperature of spin qubits contrasts with practicalities of their operation. Loss-DiVincenzo qubits, having shown the highest semiconductor-based qubit count and fidelities so far, require microwave control typically in the \SI{5}{}-\SI{40}{\giga\hertz} range \cite{Yoneda_2017,lawrie2021simultaneous,Noiri_2022_highfidelity,Xue_2022_twoqubitgate,Madzik_2022_highfidelity,Mills_2022_highfidelity}. Other encodings, such as singlet-triplet and exchange-only qubits, don't require such high frequencies but still require baseband pulses which contain spectral components on the order of \SI{1}{}-\SI{100}{\mega\hertz} \cite{nichol2017high,weinstein2023universal}. All of these signals dissipate electromagnetic radiation at the device which adds to the thermal load that must be cooled by the dilution refrigerator, depending on the quality of signal hygiene and device thermalization.

The consequences of this competition between thermal dissipation during qubit control and dilution refrigeration have arisen repeatedly in experiments over the past several years, particularly in experiments using high-frequency microwaves \cite{freer2017single,Watson_2018,Takeda_2018_freqshift,Hendrickx_2020_Ge2qubit,zwerver2022qubits,Philips_2022_6qubit,Savytskyy_2023_flipflop}. These effects, which we call ``heating effects'' for simplicity, are of increasing experimental significance due to their generally detrimental impact on qubit quality and adverse scaling with the number of control signals. The telltale evidence of the heating effect is a sizeable shift (typically of order \SI{1}{\mega\hertz}) in a spin's Larmor frequency $\omega_0 = 2\pi f_0$, which is of particular importance when attempting to drive resonant Rabi oscillations. It is remarkable that such a widespread effect, affecting semiconductor qubit devices of different architectures and material platforms, is poorly understood. Glaringly, it had not, until now, been decisively shown that the ``heating effect'' had any concrete relation to device temperature or was some other artefact of microwave dissipation.

In a single-electron spin qubit, the heating effect was shown to be overcome by using tailored pulses, with the fitted scaling of the heating effect taken as an input parameter \cite{Takeda_2018_freqshift}. For relatively small frequency shifts, another pragmatic approach is calibrating an intermediate operating frequency that always permits qubit control at the expense of some systematic error \cite{freer2017single}. In other experiments, an off-resonant microwave ``prepulse'' was used to bring the device to a steady state before carrying out experiments \cite{Watson_2018}. In a recent demonstration of a universal six-qubit silicon processor, both prepulses and ``wait times'' on the order of several \SI{100}{\micro\second} were employed \cite{Philips_2022_6qubit}. The former were required to achieve high single-qubit control fidelities, and the latter were found to be necessary to achieve satisfactory readout performance using Pauli-spin blockade (PSB) spin-to-charge conversion. None of these ad-hoc solutions are well-suited to scaling spin qubit platforms if they remain a requisite for achieving satisfactory fidelities. For example, they are incompatible with interleaved measurement and operation which will be necessary for implementing quantum error correction protocols. Furthermore, repeated observation of the heating effect implies there is missing, but important, physics in the standard Hamiltonians describing semiconductor spins. The heating effect is therefore of both fundamental and practical interest for the spin qubit community.

In this article, we present an in-depth study of the heating effect in a six-qubit silicon quantum processor. First, we outline how qubit control and measurement lead to on-chip temperature changes. Second, we make an explicit connection between device temperature and qubit frequency by directly controlling the mixing chamber temperature. This permits a phenomenological understanding of the heating effect that underpins the remaining results. Third, we illustrate how the heating effect manifests during base-temperature operation as a result of applying off-resonant microwave pulses. We show that lower-frequency baseband pulses result in frequency-shifts as well and that heat from these signals dissipates globally across the six-qubit linear array. Combining our understanding of the above effects, we are able to illustrate how the heating effect can cause non-Markovian crosstalk errors during data collection. Finally, we show how purposely increasing the mixing chamber temperature to \SI{200}{\milli\kelvin} substantially mitigates the heating effect without compromising qubit coherence or controllability.



We cannot conclusively pinpoint a microscopic mechanism for the heat-induced frequency shift. However, taking into account our observations as well as those inferred from previous publications, we are able to eliminate several possible mechanisms, and furthermore argue that an interplay of two different mechanisms may be responsible.

\section{Device control and thermalization}

\begin{figure*}[t]
    \centering
    \includegraphics[width=\textwidth]{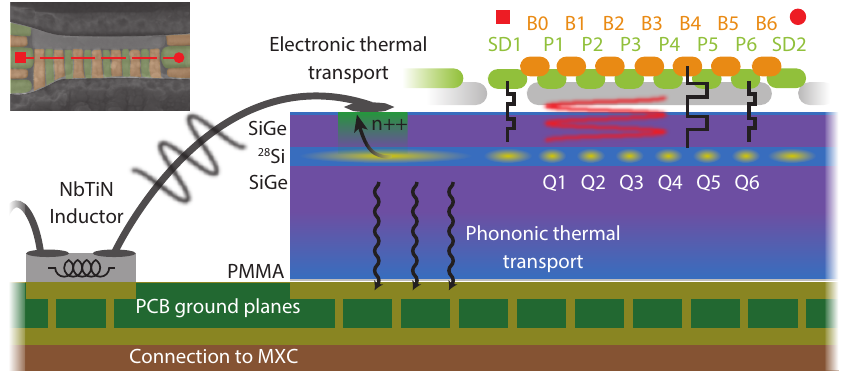}
    \caption{Cross-section of the bonded $^{28}$Si/SiGe sample (not to scale) illustrating the relevant control signals and thermal dissipation contributing to thermalization of the qubit array. The dashed line on the scanning electron microscope (SEM) image indicates the gate region labelled in the schematic. The screening gate layer (grey) carries microwave-frequency signals for driving single-qubit gates via EDSR using the spin-orbit coupling generated by the micromagnet (dark grey in SEM) stray field. The plunger gate layer (light green) carries baseband pulses for the charge sensor and for manipulating charge states within the quantum dot array. The barrier layer (orange) carries comparatively larger baseband pulses for controlling the exchange interaction used to implement two-qubit gates. Intermediate Al$_2$O$_3$ dielectric layers are not shown. Readout via reflectometry makes use of an rf signal that travels through an off-chip NbTiN high kinetic-inductance inductor and through the accumulated 2DEG via the bonded n++ Ohmic contact (the separate tank circuit used for measuring SD2 is not drawn). Phonons thermalize the crystal lattice to the dilution refrigerator through the PMMA glue between the sample and a PCB ground plane which is galvanically connected to the mixing chamber (MXC). Electrons may also carry some excess heat away from the quantum well by diffusing through the electron reservoirs, ohmic contacts, and bond wires.}
    \label{fig:1}
\end{figure*}

Extensive details on the design, control, and calibration of the six-qubit processor are already covered in \cite{Philips_2022_6qubit}. However, it is pertinent to summarize the means through which heat makes its way from the room-temperature control electronics to the sample, as well as how heat dissipates from the device to the dilution refrigerator mixing chamber. This is summarized in Fig.~\ref{fig:1}.

The sample is mounted to the gold-plated copper ground plane of a printed circuit board (PCB) with poly(methyl methacrylate) (PMMA). This is approximately a \SI{0.5}{\milli\meter} distance below the quantum well and Ti:Pd gates. Gates holding only a static DC voltage can be stripped of high frequency components using a combination of RC, Pi, and metal powder filters \cite{Mueller_2013metalpowder}. However, ac signal paths must still allow high frequencies for baseband (1-\SI{100}{\mega\hertz}) and microwave frequency (10-\SI{20}{\giga\hertz}) control. In the absence of any signal, thermal fluctuations along the attenuated path will give rise to photons carrying heat toward the device. When ac signals are applied, radiation incident on the device is expected to cause additional Joule heating through a combination of conductive and dielectric losses.

In quantum dot devices operated in a dilution refrigerator, it is a general rule that the minimum equilibrium electron temperature $T_\textrm{e}$ of the accumulated electron reservoirs (2DEG) will be higher than the minimum mixing chamber temperature $T_\textrm{MXC}$ \cite{Kautz_1993_heating}. In our setup, $T_\textrm{MXC}$ can reach a base temperature of \SI{12}{\milli\kelvin}. The electron temperature floor, in contrast, is measured to be about $T_\textrm{e}\approx\SI{180}{\milli\kelvin}$ (see Appendix~\ref{sec:Elec_temp}). We believe that the local environment of the quantum dots (for instance the lattice and the bath of nearby fluctuators) reaches thermal equilibrium below the electron temperature floor due to measurable changes in the spin properties as the mixing chamber temperature is reduced below \SI{180}{\milli\kelvin} (e.g. see Fig.~\ref{fig:2}).

When the lattice temperature is below $T_\textrm{e}$, cooling of the quantum dot region to the mixing chamber is only mediated by phonons which must pass through the substrate, the PMMA ``glue'' (which is an insulating material), the PCB, the cold finger to which it is mounted, and ultimately the mixing chamber itself. The rate at which phonons are able to carry heat from the quantum dot region to the mixing chamber will depend on their temperature difference and the thermal conductance along the path connecting them. When the local environment reaches a temperature above $T_\textrm{e}$, electronic heat transport also plays a role in the cooling of the device \cite{Gustafsson_2013_thermal}. As illustrated in Fig.~\ref{fig:1}, the doped Ohmic contacts (two of which are in series with the rf reflectometry readout circuit) provide a conductive route for heat to dissipate through the wire bonds electrically connecting the sample to the PCB and the thermal anchoring of the wires at the mixing chamber.

In order to implement universal control across the six-qubit array, a variety of baseband pulses and high-frequency signals are required. Initialization and measurement of spins are performed with a combination of Pauli spin blockade (PSB) readout for spin-to-charge conversion as well as rf reflectometry to detect the change in resistance of a nearby single-electron transistor (SET). Two sets of signals are required. First, a combination of baseband pulses on plunger and barrier gates are used to tune the electrostatic landscape from the symmetric operation point to the interdot transition where PSB readout occurs \cite{reed2016symmetric}. Since the SET is typically tuned to Coulomb blockade during qubit operation to avoid unnecessary dissipation, baseband pulses are also used to tune the SET to a sensitive Coulomb peak flank while the measurement takes place. The baseband pulses consist of rise times in the range of 10-\SI{100}{\nano\second} and magnitudes on the order of \SI{10}{\milli\volt} at the quantum dot gate electrodes. Second, rf signals are applied through the electron reservoirs to measure changes in the sensing dot resistance. These tones are at \SI{88}{\mega\hertz} and \SI{94}{\mega\hertz} for the two charge sensors and have nominal powers of \SI{-30}{\dBm} and \SI{-33}{\dBm} respectively for the duration of the measurement, which is on the order of \SI{10}{\micro\second}.

Single-qubit rotations are performed with electric-dipole spin resonance (EDSR) and require microwave frequencies in the range of 10-\SI{20}{\giga\hertz}, depending on the magnitude of the qubits' Zeeman energy. Our system can perform 90-180 degree rotations in the Bloch sphere with typical gate times of 100-\SI{200}{\nano\second}. The nominal applied power of roughly \SI{-30}{\dBm}, is expected to produce an ac electric field amplitude of order \SI{100}{}-\SI{1000}{\volt/\meter} at the quantum dot array for EDSR of this frequency. Two-qubit operations require baseband pulses predominantly on the barrier gates between quantum dots. A typical CZ gate can be performed using a \SI{100}{\nano\second} baseband pulse. Due to the relatively smaller lever arm of the barrier gates, a larger pulse magnitude of \SI{100}{}-\SI{200}{\milli\volt} is required.


The heating caused by the combination of these operations is in competition with the rate of thermal dissipation away from the quantum well to the mixing chamber heat sink. Due to the high frequency, single-qubit control with microwaves, whether by EDSR or electron spin resonanace (ESR), has been the dominant origin of the heating effect in recent experiments. However, we will show that it is not the only relevant mechanism, as repeated baseband pulses will also lead to measurable frequency shifts. 

\section{Temperature Dependence of Qubit Frequencies}

\begin{figure}[t]
    \centering
    \includegraphics[width=\linewidth]{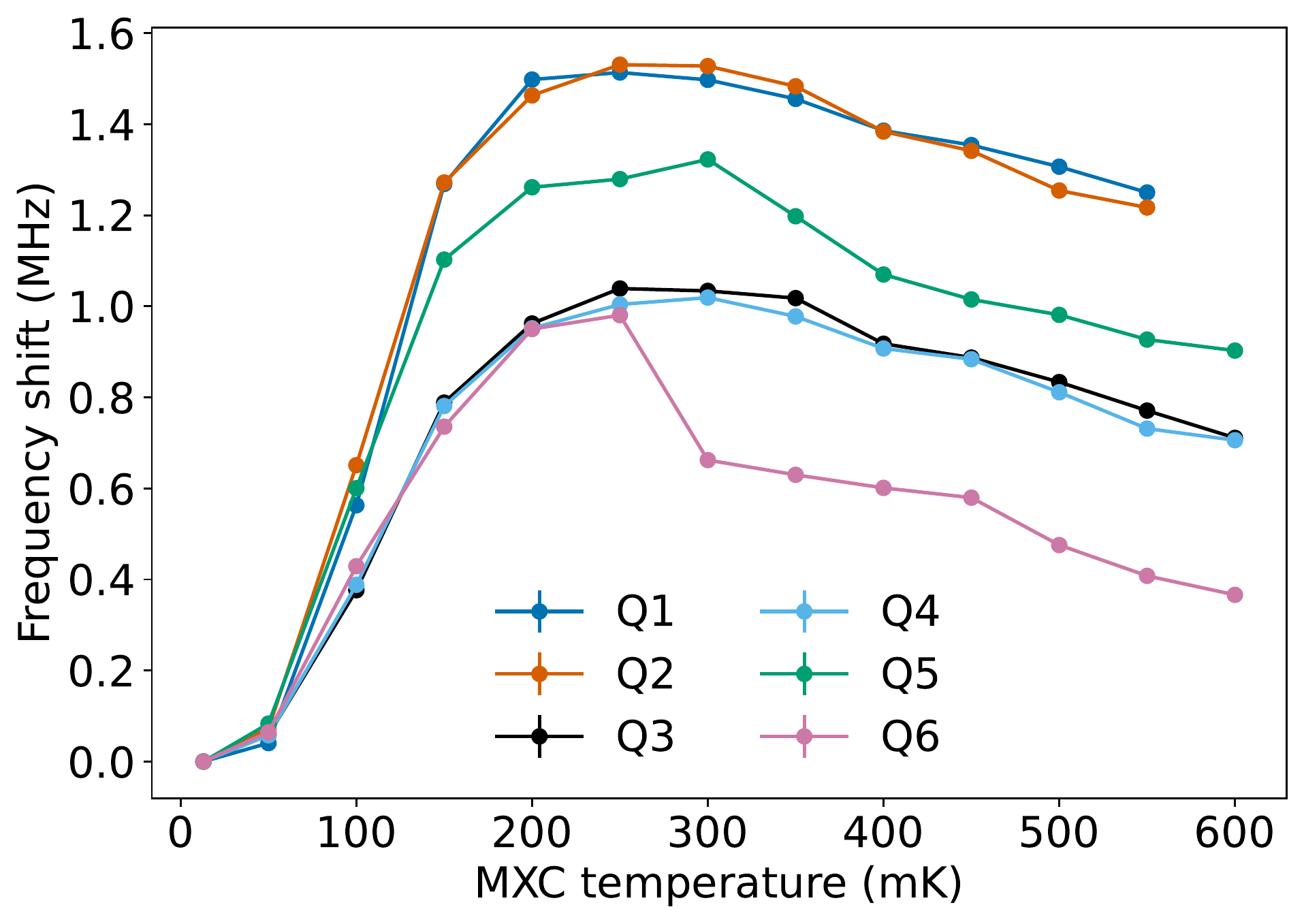}
    \caption{Larmor frequency shifts of all six qubits as a function of mixing chamber temperature measured via a Ramsey-type pulse sequence. The frequency shift is recorded relative to a measurement taken at \SI{13}{\milli\kelvin}. The device was allowed to thermalize for several minutes before each data point was collected and no gradient in the separate mixing chamber thermometer was observed. DC voltages were unchanged throughout the experiment, which took course over the period of about 12 hours. At \SI{300}{\milli\kelvin}, an environmental charge jump close to qubits 5 and 6 was observed causing a discrete shift in frequency. The electrostatic nature of this jump is inferred as it also shifted the PSB voltage window where spin-to-charge conversion takes place during qubit readout.}
    \label{fig:2}
\end{figure}

To confirm that the heating effect can be related to the temperature of the device, we make use of a PID controller to manipulate the temperature of the dilution refrigerator mixing chamber from base temperature to \SI{600}{\milli\kelvin}, as measured by a thermometer mounted to the mixing chamber, and let the device thermalize at each set point for several minutes to achieve a steady-state condition. Fig.~\ref{fig:2} summarizes the results of performing a Ramsey measurement at each temperature, using the frequency difference between the fitted free-induction decay and a known ``virtual'' detuning to infer a shift in the qubit's Larmor frequency with an accuracy on the order of \SI{10}{\kilo\hertz}. The range of temperatures explored is limited by temperature dependence of the rf reflectometry readout, which is discussed in depth in Appendix~\ref{sec:readout}.

Three features of the observed temperature dependence are particularly noteworthy. First, all six Larmor frequencies shift to higher frequencies. We may note that previously published results, making use of microwave pulses to probe the heating effect as opposed to directly varying the mixing chamber temperature, report both positive frequency shifts \cite{Watson_2018, Xue_2019_Benchmarking, freer2017single, Savytskyy_2023_flipflop} and negative frequency shifts \cite{zwerver2022qubits, Takeda_2018_freqshift} (the latter reports both). Second, all six frequencies shift by a similar magnitude of roughly \SI{1}{\mega\hertz}. This, again, is similar to previously reported pulse-induced shifts which range from tens of \SI{}{\kilo\hertz} to a few \SI{}{\mega\hertz}. Third, all shifts exhibit a striking non-monotonic trend. These observations suggest physics that acts globally across the qubit array with some degree of variation between quantum dots.

We are unaware of any theoretical precedent for describing how the qubit frequencies shift with temperature. As all qubit frequencies are in the \SI{16}{\giga\hertz} band, a shift of \SI{1}{\mega\hertz} constitutes less than a 0.01\% change. Since the $g$-factor of the confined electron spins and the total magnetic field at the quantum dot are not known to this precision, there are many open possibilities regarding the microscopic origin of this effect. We save an in-depth discussion for Section~\ref{sec:discussion}.

It is worth noting that the ``cold'' qubit frequency, which we use as the origin for measuring any relative frequency shifts, is somewhat difficult to probe. As illustrated in Section~\ref{sec:cooling}, the device thermalizes very slowly at base temperature, and therefore the brief microwaves required to perform the Ramsey experiment cause a systematic self-heating. While long wait times of typically 0.5-\SI{1}{\milli\second} are inserted between each measurement to mitigate this effect, we find it is still insufficient to keep the device at equilibrium with the \SI{12}{\milli\kelvin} mixing chamber. Rather, the sparse sequence of $X_{90}$ pulses may be enough to raise the average device temperature closer to \SI{50}{\milli\kelvin} as suggested by the adjacent data point. As will be illustrated in Section~\ref{sec:mitigation}, this systematic error becomes negligible as $T_\textrm{MXC}\gtrapprox\SI{150}{\milli\kelvin}$ and the heat capacity of the heterostructure increases.

\FloatBarrier

\section{The heating effect}

\begin{figure*}[t]
    \centering
    \includegraphics[width=0.99\textwidth]{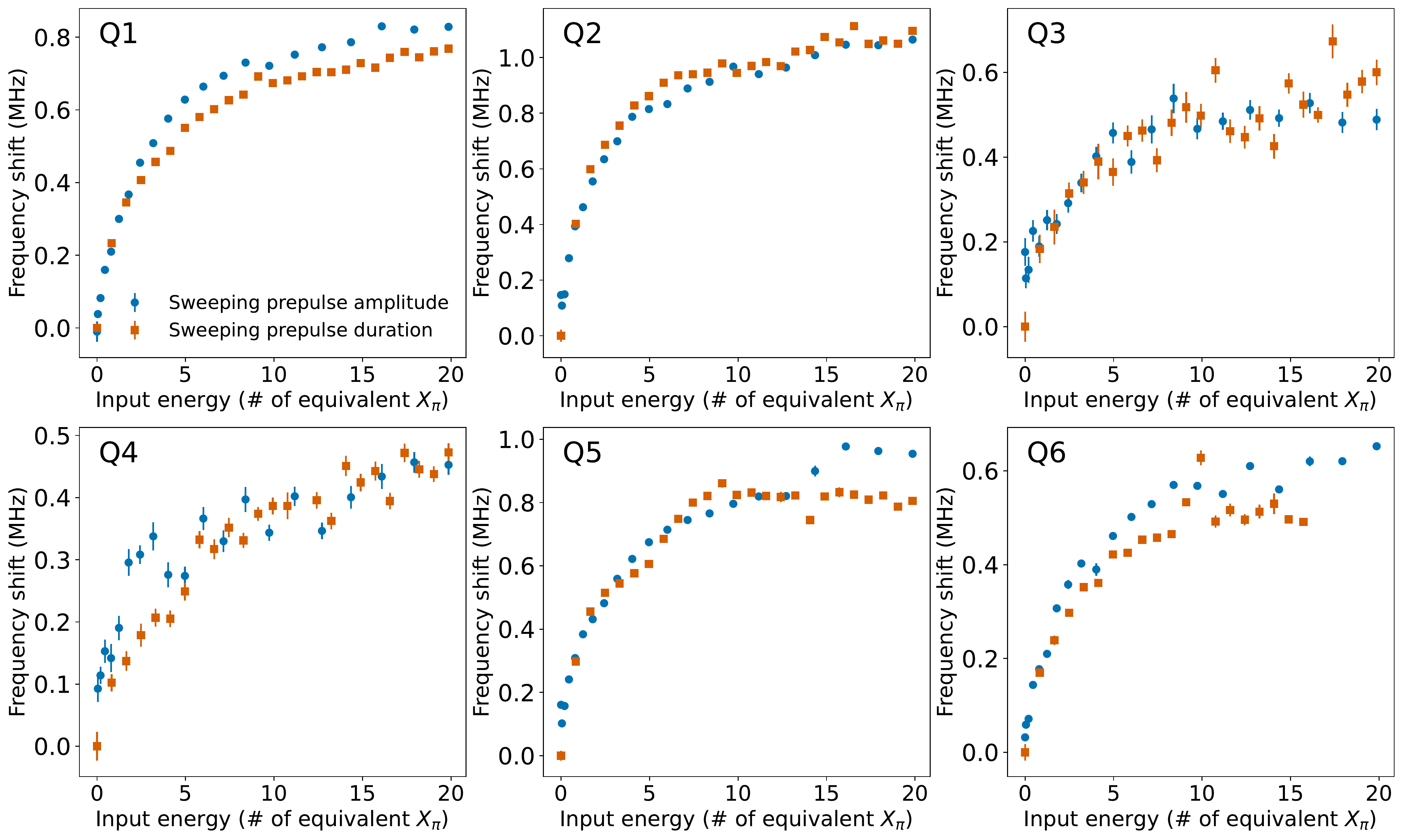}
    \caption{Frequency shifts at \SI{20}{\milli\kelvin} mixing chamber temperature measured after applying an off-resonant pulse of various duration and amplitude. For a swept prepulse amplitude, a fixed duration of \SI{2}{\micro\second} was used. For a swept prepulse duration, a fixed amplitude corresponding to $f_\textrm{Rabi}\approx\SI{1.3}{\mega\hertz}$ was used. Prepulse energy is scaled to ``equivalent-gate'' energy by averaging the relative pulse amplitudes and durations required to drive EDSR on all six qubits in the array with $f_\textrm{Rabi}=\SI{2}{\mega\hertz}$. The precise electric field amplitude of the prepulse at the quantum well is unknown, but expected to be on the order of \SI{1000}{\volt/\meter} according to the estimated dot sizes, micromagnet gradient, and observed Rabi frequencies. A \SI{0.5}{\milli\second} wait time was implemented between all shots. Frequency shifts were measured by fitting a chevron-pattern line-cut after the off-resonant pulse was turned off.}
    \label{fig:4}
\end{figure*}

\subsection{Microwave induced heating}

We now document the manifestation of the heating effect in all six qubits due to off-resonant microwave pulses while the mixing chamber is set to \SI{20}{\milli\kelvin}. Fig.~\ref{fig:4} illustrates two complementary microwave pulse experiments carried out at \SI{20}{\milli\kelvin}. After initialization of the qubit under test, an off-resonant microwave burst at a constant frequency of \SI{16}{\giga\hertz} is applied. This frequency is selected since it is approximately \SI{100}{\mega\hertz} below the lowest qubit Larmor frequency and does not give rise to any significant coherent driving. The off-resonant frequency is held constant for all qubit experiments since different frequencies may transmit and dissipate in the device differently. To measure the pulse-induced frequency shift, we subsequently scan the frequency of a second microwave pulse, calibrated in length to result in a 180 degree rotation. The resonance peak may then be fit to Rabi's formula to estimate the Larmor frequency.

Since our measurement ``probe'', the resonant microwave pulse in combination with an rf reflectometry pulse, is also expected to deposit heat in the system, it is necessary to scrutinize to what extent our results are affected by ``self-heating'' of the measurement procedure itself. This will be discussed in further detail in Section~\ref{sec:cooling}. For now, we emphasize that the measurement pulse is shorter than most of the off-resonant pulses used during the scan, and therefore the self-heating should only account for a small fraction of the total effect. Furthermore, it is a constant for all experiments, so any difference in results can be attributed to the varied off-resonant pulse. To ensure the independence of all measurement shots, a long wait time between sequences of \SI{500}{\micro\second} is used to allow most accrued heat to dissipate before the next experiment begins.

Fig~\ref{fig:4} shows the Larmor frequency shifts as a result of sweeping the off-resonant pulse amplitude and duration, respectively. When the amplitude and duration scales are converted to equivalent gate-operation energies, the severity of the heating effect becomes evident \footnote{For this conversion we use an amplitude equal to the average of the amplitudes used to drive each qubit independently with $f_\mathrm{Rabi}=\SI{2}{\mega\hertz}$}. After only approximately 10 $X_\pi$ gates worth of energy, each qubit's Larmor frequency has shifted by \SI{0.3}{}-\SI{1}{\mega\hertz}. A microwave drive originally calibrated to a qubit's ``cold'' frequency is no longer able to drive complete spin rotations.

To plot the measured frequency shifts on a common energetic axis, a Joule heating model is assumed \cite{Pozar_2012}:

\begin{equation}
    E_\mathrm{Joule} = \int_t\mathrm{d}t\int_V \mathrm{d}\mathbf{r}\frac{\sigma(\mathbf{r}) + \omega\epsilon''(\mathbf{r})}{2}|E_\mathrm{ac}|^2.
    \label{eq:joule}
\end{equation}

\noindent In this model, the electromagnetic radiation propagating from the on-chip EDSR antenna will dissipate in the device due to a combination of conductive losses ($\sigma$), dielectric losses ($\epsilon''$ is the complex component of the dielectric function $\epsilon = \epsilon' - j\epsilon''$) throughout the volume $V$ penetrated by the fields. Magnetic loss due to complex permeability is neglected for simplicity. Due to the many different materials on the device, and the uncertain volume in which dissipation occurs, the exact integration is not possible to perform. However, by making the simplifying assumption that the electric field amplitude profile $f(\mathbf{r})$, where $E_\mathrm{ac}(\mathbf{r},t) = |E_\mathrm{ac}|\cos(\omega t)f(\mathbf{r})$, is the same for all prepulses and that all material parameters are time-independent, the volume integration can be treated as a constant in all cases, and the deposited energy is estimated to scale quadratically with the drive amplitude and linearly with the drive duration. With this model, we find that the frequency shift is set by the energy deposited in the device. This, in addition to the observation that all qubit frequencies once again shift upwards, suggests that the dissipation of the applied microwave signal is consistent with the observed temperature dependence of the Larmor frequency. Although the presence of microwaves may enable other possible frequency-shift mechanisms, such as directly via the ac Stark effect or indirectly via coupling to environmental degrees of freedom, the temperature dependence is dominant.

\subsection{Baseband induced heating}

\begin{figure*}[t]
    \centering
    \includegraphics[width=\textwidth]{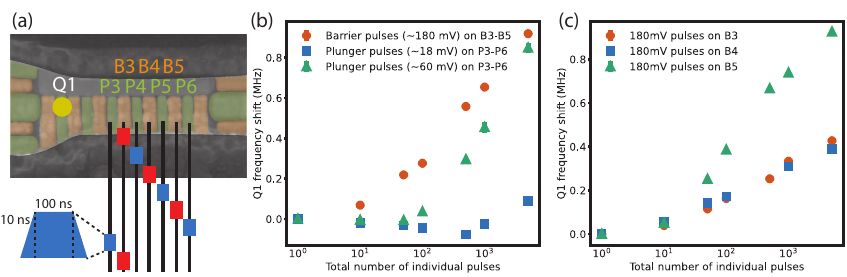}
    \caption{(a) Illustration of the serially applied pulses used to induce baseband signal heating. The frequency shift of Qubit 1 can be correlated with the degree of nonlocal heating. (b) Frequency shift of Qubit 1, measured via a Ramsey experiment, as a result of various combinations of applied pulses. All pulses are \SI{100}{\nano\second} in duration with \SI{10}{\nano\second} ramps. Pulses are applied back-to-back with no intermediate wait between ramp-down and ramp-up. As there is effectively zero lever arm between the gates used and the quantum dot hosting Qubit 1, any shift is expected to arise due to heating and not due to a net shift in the micromagnet gradient. Data is normalized to the measurement with zero baseband pulses. The negative shift for \SI{18}{\milli\volt} plunger pulses likely arises from some low-frequency drift in qubit frequency which can be as much as tens of \SI{}{\kilo\hertz} during hour-long measurements such as this one. (c) Frequency shift of Qubit 1 as a result in repeated pulses on individual barrier gates. Pulses sent to B5 appear to dissipate the most heat, causing a greater frequency shift in Qubit 1 than for the same nominal pulses applied to B3 and B4.}
    \label{fig:5}
\end{figure*}

Although high frequency microwaves will dissipate more energy into the device than lower-frequency baseband pulses, it is prudent to check the effect of these necessary signals. We employ Qubit 1 as an on-chip ``thermometer'' and send baseband pulses to plunger and barrier gates on the opposite side of the device as illustrated in Fig.~\ref{fig:5}(a). All of the participating gates have effectively zero dc lever-arm with respect to the quantum dot containing Qubit 1.

In Fig.~\ref{fig:5}(b), we apply a variety of pulse amplitudes mimicking standard device operations. These operations are applied one at a time and cycle through the relevant gates from B3 to P6. Small plunger pulses (\SI{10}{\milli\volt}) mimic what is required to detune dots between the operation point and the Pauli spin blockade readout point. These contribute little heat to the device even after thousands of operations. Large plunger pulses (several tens of \SI{}{\milli\volt}) replicate the amplitude with which one may scan gates while doing charge state tuning. The larger voltage ramps do cause heating after hundreds of pulses, and are consistent with our observation that the mixing chamber temperature increases by tens of \SI{}{\milli\kelvin} when doing ``video mode'' charge state tuning with continuous rf reflectometry, despite the fact that there is no nominal attenuation in the high-frequency line at this temperature stage.

Barrier pulses cause the largest baseband heating effect as their relevant working mode - turning on and off the exchange interaction - requires the largest voltage pulses with fast rise times. Tens of barrier pulses have a substantial impact on qubit frequency due to the heating effect. This may also impact qubit encodings other than Loss-DiVincenzo qubits; exchange-only qubits already require 10-100 such pulses for small algorithms (although, in the absence of micromagnets, the effects may be less severe when operating at very low magnetic fields).

A sublinear relation between the number of pulses and the total frequency shift is observed, similar to what is observed when heating is induced by microwave prepulses and a saturation effect is seen as more energy is deposited into the device. We expect that, due to the tens-to-hundreds of microseconds of baseband pulses required to induce the frequency shifts, there will also be an interplay of device cooling and heating contributing to the net frequency shift measured.

By separating the gates receiving the baseband pulses, we can estimate the locality of the heating effect. In Fig.~\ref{fig:5}(c), we measure the frequency shift of Qubit 1 after a series of pulses on individual barrier gates. Although pulses on all three tested gates have an effect, it is curious that the gate farthest from Qubit 1, B5, imparts the largest shift. This suggests that the heating effect acts globally across the six qubit array. Larger qubit arrays or bespoke test structures would be required to experimentally determine the length scale of dissipation.

\subsection{The cooling effect}
\label{sec:cooling}

\begin{figure}
    \centering
    \includegraphics[width=\linewidth]{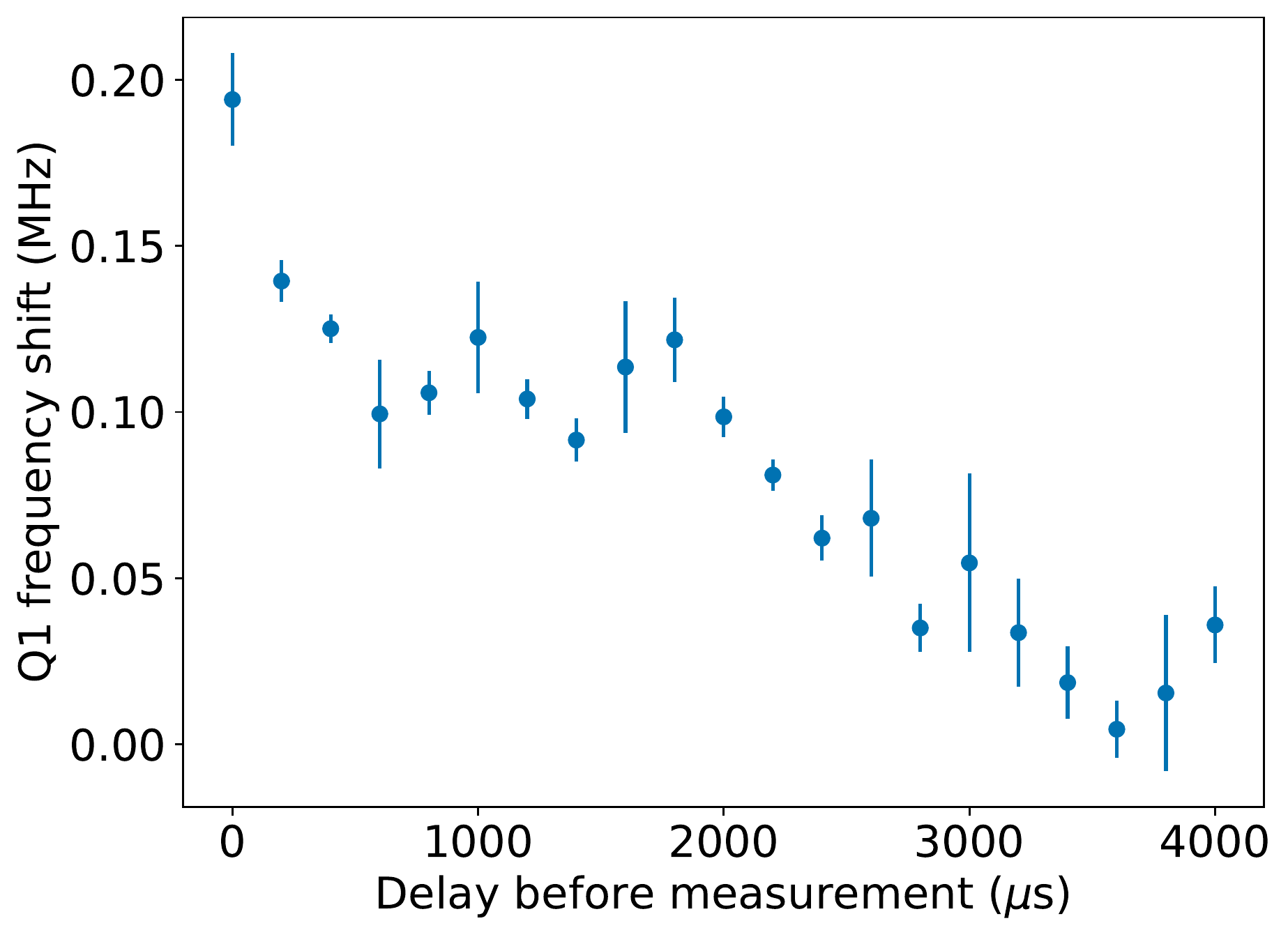}
    \caption{The cooling effect after applying a prepulse with approximately 3 $X_\pi$ gates of energy at a mixing chamber temperature of \SI{50}{\milli\kelvin}. The horizontal axis measures the time between the off-resonant prepulse and the qubit frequency measurement. A wait time of \SI{5}{\milli\second} is used between measurements to allow complete inter-shot cooling. The frequency shift is compared to the ``true'' \SI{50}{\milli\kelvin} qubit frequency measured with a Ramsey experiment using the same long wait time}
    \label{fig:6}
\end{figure}

In analogy to the qubits showing a heating effect, they also exhibit a cooling effect. This is the frequency shift that occurs after a qubit has been warmed up by a microwave pulse as it equilibrates with its environment. Around base temperature operation, the cooling time is very slow with respect to the total time of a typical measurement cycle. The former has been measured to be milliseconds, and the latter is on the order of tens or hundreds of microseconds. Fig.~\ref{fig:6} illustrates one instance of Qubit 2 cooling down after an off-resonant microwave burst caused a \SI{200}{\kilo\hertz} heat-induced frequency shift. The cooling time-scale is not substantially influenced by the flow rate of the $^3$He/$^4$He mixture through the dilution refrigerator, suggesting that the cooling is limited by thermal dissipation to the mixing chamber.

We may also comment that the cooling effect, unlike the heating effect, showed some variation over the total duration of the experiments presented here. The cooling time scale varied by an order-unity factor that seemed to depend somewhat on the tuning history of the device. This perhaps indicates that particular electrostatic configurations are more amenable to transporting thermal energy away from the qubit array, but we were unable to study this systematically. It is also challenging to measure cooling timescales at higher device temperatures, as the heating effect becomes negligible. We discuss this further in the next section.

\section{Crosstalk and mitigation}
\label{sec:mitigation}

\subsection{A phenomenological picture}

\begin{figure*}
    \centering
    \includegraphics[width=0.8\textwidth]{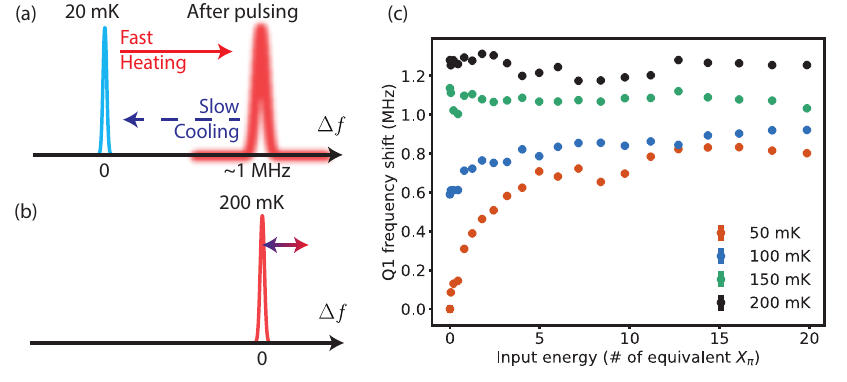}
    \caption{(a) Phenomenological illustration of a qubit resonance shifting in response to signal dissipation at \SI{20}{\milli\kelvin}. The Larmor frequency will quickly shift upwards to an approximate steady state when the device temperature increases. Slow thermalization causes the resonance to drift back to its initial state. Because of this interplay, deterministically pulsing to a ``warm'' steady-state from base-temperature operation is difficult. (b) When the mixing chamber and device are kept at \SI{200}{\milli\kelvin}, the resonance is much more robust to the smaller thermal fluctuations caused by signal dissipation. (c) Frequency shifts of Qubit 1 observed at various mixing chamber temperatures as a result of applying an off-resonant prepulse of varying energy and measuring the shift in the resulting chevron pattern. The energy scale is the same as that used in Fig.~\ref{fig:4}. The absolute increase in frequency with increasing temperature is directly comparable with the temperature dependence presented in Fig.~\ref{fig:2}. Other qubits in the array exhibit the same phenomenon.}
    \label{fig:7}
\end{figure*}

With the observed dependence of the qubits' Larmor frequencies on the device temperature, we can introduce a phenomenological picture of the heating effect, illustrated in Fig.~\ref{fig:7}(a), from which the following results can be qualitatively understood. The Larmor frequency of a particular qubit is well defined when the device is at thermal equilibrium with the mixing chamber. When high-frequency signals are applied, the dissipated heat will shift the Larmor frequency upwards. The shift occurs on a time scale faster than the \SI{100}{\nano\second} of coherent microwave driving necessary to measure the frequency shift. In contrast, the cooling time scale is on the order of milliseconds for base temperature operation but much faster at higher temperatures as more efficient thermal pathways are made available. Due to competition between heat dissipation and cooling power, the resonance condition moves on a timescale similar to individual measurements. This makes it difficult to calibrate a microwave prepulse to deterministically shift the Larmor frequency during a series of measurements. The resulting frequency shifts, although systematic in nature, are not easily predictable and behave as an additional noise source. A qubit frequency that varies between experiments will negatively impact the $T_2$ coherence properties. An overall decrease in $T_2^*$ when employing prepulsing at base-temperature operation was previously documented for all qubits in this device \cite{Philips_2022_6qubit}.


A systematic detuning error also limits the fidelities with which spins can be resonantly manipulated. As a concrete example, the infidelity of a 180 degree resonant spin flip (in the single-qubit Hilbert space) scales as $1-F \approx \frac{2}{3}\left(\frac{\Delta^2}{\Omega^2 + \Delta^2}\right)$ where $\Delta = \omega_\textrm{MW} - \omega_0$ is the detuning between the microwave drive frequency ($\omega_\textrm{MW}$) and the target qubit Larmor frequency ($\omega_0$) and $\Omega$ is the effective drive amplitude regardless of whether ESR or EDSR is employed. When $\Delta = 0$, $\Omega/2\pi$ can be identified one-to-one with the Rabi frequency $f_\textrm{Rabi}$. For few-\SI{}{\mega\hertz} Rabi frequencies often used in silicon spin qubit systems, a maximum detuning of about \SI{100}{\kilo\hertz} is required to achieve at least 99.9\% fidelities. This range can be compared to the typical heating effect Larmor frequency shift of \SI{1}{\mega\hertz}. Without bespoke calibration techniques \cite{Takeda_2018_freqshift,Philips_2022_6qubit}, high-fidelity operation is compromised. 

It is also worth noting that a variable detuning is also deleterious for idling qubits. Since the microwave frequency also defines the ``rotating frame'' for each qubit wherein Pauli-Z operations can take place through phase updates, unaccounted dynamics in the Larmor frequency will result in unknown phase accumulation. Therefore, even though a particular qubit may be spectrally well-separated from its resonantly-driven neighbour, the heating effect opens up a large channel for crosstalk errors that must also be systematically calibrated.

The spin qubit literature almost exclusively treats the Larmor frequency $\omega_0$ as a constant, with some added random noise distribution giving rise to $T_2$-type decoherence as a result of charge noise and contact hyperfine interaction with nearby nuclear spins. This modeling breaks down in the presence of the heating effect, as $\omega_0$ undergoes systematic and contextual shifts that are not accurately captured by a Markovian environment. These shifts can far exceed the intrinsic line width of the qubit resonances, which is estimated to be on the order of \SI{10}{\kilo\hertz} for the best published devices. Due to the fundamental role of $\omega_0$ in spin-based quantum computing, the heating effect has impact beyond single-qubit control and also interferes in optimizing two-qubit gate fidelities, for example.

\subsection{The benefit of warmer operation}

Next, we investigate whether the harmful frequency shifts from microwave excitation may be reduced by directly raising the device temperature via the dilution refrigerator mixing temperature as opposed to using microwave prepulsing. A PID controller connected to an off-chip resistance thermometer is used to regulate the sample temperature. The Pauli spin blockade readout mechanism, unlike energy-selective readout to an electron reservoir, is not compromised by operating at higher temperatures, and it is already established that universal quantum logic is achievable with spin qubits at these temperatures \cite{yang2019hotsilicon,Petit_2020hot}.

Fig.~\ref{fig:7}(c) illustrates the effect of an off-resonant burst on two qubits at different mixing chamber temperatures. Curiously, the heating effect becomes smaller at a working temperature of \SI{100}{\milli\kelvin} and seems to vanish completely at temperatures above \SI{150}{\milli\kelvin}. Given the non-monotonic temperature dependence of the qubit frequency, it is natural to ask whether this is the effect of operation close to the temperature ``sweet spot''. The answer does not seem to be this simple. Since a \SI{100}{\milli\kelvin} operating point sits at a steep slope of the Larmor frequency shift trend (see Fig.~\ref{fig:2}), this does not explain why off-resonant pulses induce a smaller frequency shift here. Furthermore, the absence of a frequency shift at \SI{150}{\milli\kelvin} would be peculiar since the derivative is nonzero here as well.

Two possibilities are consistent with this observation. The first is that excess heat is able to dissipate more quickly as the device temperature is increased. In addition to the dilution refrigerator providing more cooling power at higher temperatures, electronic thermal transport may be able to carry excess heat away from the qubit array more efficiently than phononic thermal transport when the device temperature is above the electron temperature floor. The second possibility is that the specific heat of the materials in the quantum dot region increases. It follows from the Debye model that the specific heat of the crystalline heterostructure should increase as $T^3$. Therefore, the heat deposited from signal dissipation will cause a much less significant temperature rise at $T\gtrapprox\SI{150}{\milli\kelvin}$ versus $T\approx\SI{15}{\milli\kelvin}$. This picture is simply illustrated in Fig.~\ref{fig:7}(b), where, in contrast to microwave prepulsing, directly raising the temperature of the device via the mixing chamber leads to much less Larmor frequency shifting and, counter-intuitively, spins which are easier to control. 

\begin{figure}
    \centering
    \includegraphics[width=\linewidth]{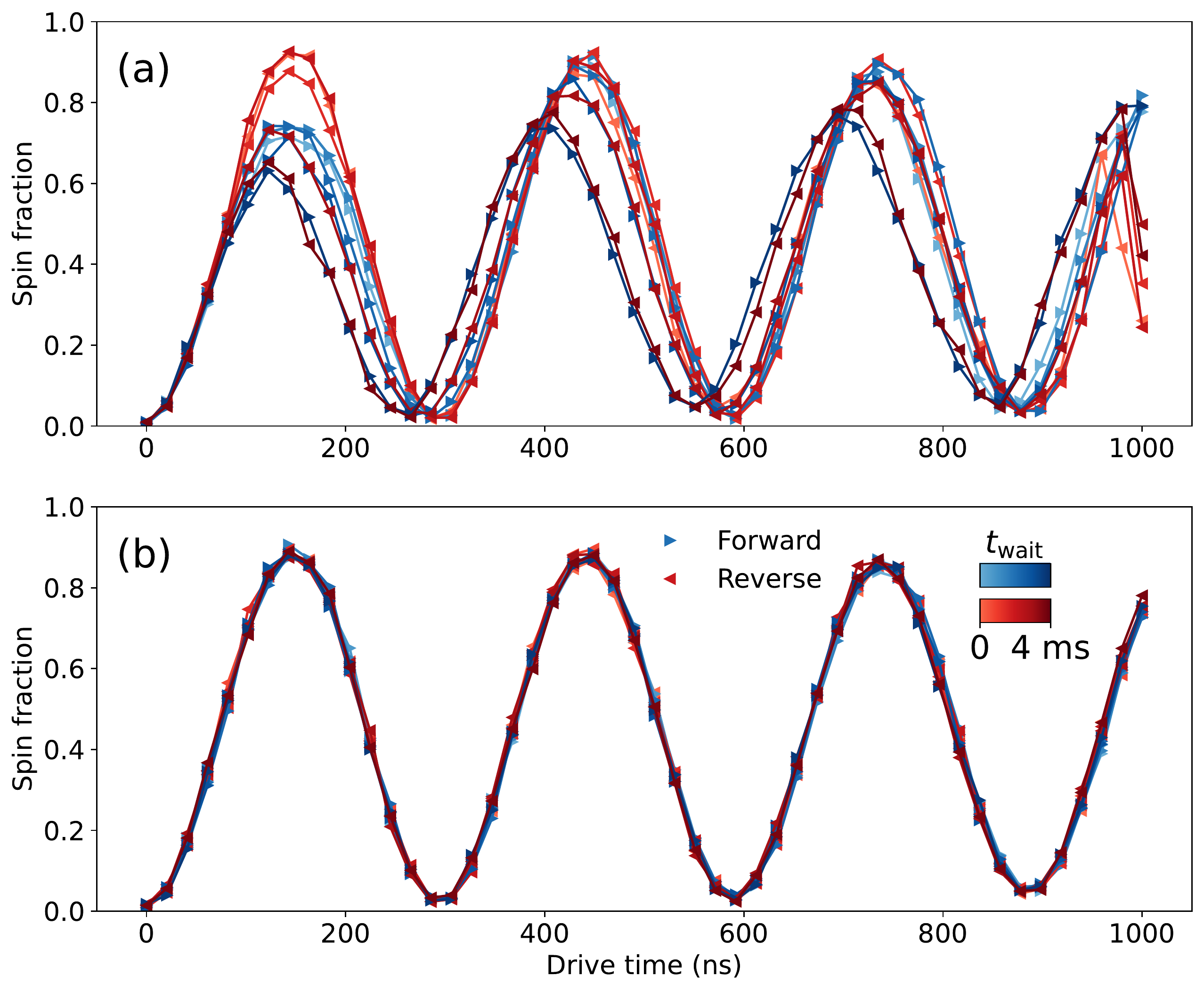}
    \caption{Rabi oscillations of Qubit 1 collected at a mixing chamber temperature of (a) \SI{12}{\milli\kelvin} and (b) \SI{200}{\milli\kelvin}. For each set of measurements, the qubit frequency is calibrated using a ``warm'' chevron pattern. The wait time $t_\textrm{wait}$ is added between measurement and initialization sequences, while ``Forward'' and ``Reverse'' scans collect data from \SI{0}{} to \SI{1000}{\nano\second} and \SI{1000}{} to \SI{0}{\nano\second} respectively. 2000 shots are averaged for each datapoint.}
    \label{fig:8}
\end{figure}

To illustrate the benefit of qubit operation at warm temperatures, we focus on a particular qubit and compare Rabi oscillations at \SI{12}{\milli\kelvin} and \SI{200}{\milli\kelvin} mixing chamber temperatures. The latter temperature is not uniquely special; it is high enough to mostly escape the heating effect while low enough such that readout visibility is not compromised. Fig.~\ref{fig:8} compares the two cases. At each temperature, the rotating frame is calibrated to the qubit's approximate ``warm'' Larmor frequency. Each data point consists of 2000 consecutive shots and experiments are run twice: from \SI{0}{}-\SI{1}{\micro\second} and in reverse. Furthermore we vary the wait times used between experimental shots from zero (which still includes readout integration times and necessary delays for the electronics totalling tens of \SI{}{\micro\second}) to \SI{4}{\milli\second}.

Common intuition built off of a simple qubit Hamiltonian would suggest that all collected statistics would be identical. In fact this is what we see at the \SI{200}{\milli\kelvin} operating point. At base temperature, however, the Rabi oscillations show a great degree of variation due to the interaction between the heating effect, the cooling effect, and the order and speed with which data is collected. The cumulative effect is a striking illustration of non-Markovian behaviour, where resonant spin control becomes highly contextual on the history of the device operation. Higher temperature operation no longer appears only convenient, but essential, for high-quality quantum computing with microwave-controlled qubits in the presence of the heating effect.

To further illustrate the utility of \SI{200}{\milli\kelvin} operation, we perform single-qubit randomized benchmarking on all qubits and find all average single-qubit gate fidelities are in the range of 99.8\%-99.9\%. These values are directly comparable to previous values obtained at base temperature \cite{Philips_2022_6qubit}, but require no off-resonant prepulses, pulse-shaping, or other ad-hoc calibrations to achieve; fitting the chevron pattern to calibrate the Larmor frequency along with a fit of Rabi oscillations to set the gate time is sufficient to reach this level of performance. We expect these values could be further improved with an optimized $f_\textrm{Rabi}$ and sophisticated pulse shaping. Furthermore, the phase picked up by idling qubits during the off-resonant drive scales linearly with the duration of the pulse making crosstalk calibration more straightforward. Lastly, we comment that qubit coherence times are negligibly affected by working at the warmer temperature, as measurements of $T_2^*$ shown in Appendix~\ref{sec:coherence} illustrate. Additional effects of temperature on qubit coherence properties and more for this device are extensively explored in a separate manuscript.


\section{Origin of the temperature dependence}
\label{sec:discussion}

The physics underpinning the heating effect is the explicit temperature dependence of the qubits' Larmor frequencies on device temperature. It is difficult to identify the precise microscopic origin of this effect, but we may comment on the plausibility of a number of potential mechanisms in light of the experimental data presented here. Each qubit frequency can be expressed as $\omega_i = g_i\mu_B B_\textrm{tot,i}/\hbar$, where $g_i$ is the electron spin $g$-factor, $B_\textrm{tot,i}$ is the total magnetic field at the quantum dot location, $\mu_B$ is the Bohr magneton, and $\hbar$ is the reduced Planck constant. In general, both $g_i$ and $B_\textrm{tot,i}$ may exhibit temperature dependence directly or indirectly.

In principle, it is possible to experimentally isolate $g_i$ from $B_\textrm{tot,i}$ in our setup by measuring the qubit frequency as a function of the total magnetic field. However, we are interested in a shift with a relative magnitude of 0.01\%. Measuring the $g$-factor with this degree of accuracy requires knowing the total magnetic field at the quantum dot location with high precision, and this is impractical given the presence of an on-chip micromagnet.

In Appendix~\ref{sec:Stark}, we include temperature-dependent Larmor frequency measurements taken at three different external magnetic field settings. A temperature change of this magnitude should have no effect on the field provided from the external magnet as the superconducting solenoid is far from the mixing chamber and carries a persistent superconducting current. We do observe a larger frequency shift at larger external field values, which is consistent with the observation of other work suggesting the microwave-induced frequency shift originates with the $g$-factor \cite{freer2017single,zwerver2022qubits,Savytskyy_2023_flipflop}. However, while we operate in a regime where the micromagnet magnetization is effectively saturated (i.e. the qubit frequency scales linearly with the external field), we cannot guarantee this to within 0.01\% accuracy and thus cannot rule out any temperature-dependence in the micromagnet stray field.

Residual $^{29}$Si in the purified quantum well and $^{73}$Ge nuclear spins in the SiGe buffer contribute to an effective Overhauser field that may exhibit temperature dependence through Boltzmann statistics. Both isotopes have negative gyromagnetic ratios of \SI{-8.46}{\mega\hertz/\tesla} and \SI{-1.49}{\mega\hertz/\tesla}. The energy splitting of these nuclear spin states is on the order of \SI{0.01}{\micro\eV}, meaning that at base temperature where $k_BT\approx\SI{1}{\micro\eV}$ there is a very slight nuclear polarization opposing the external field. As the temperature is increased, the magnitude of the Overhauser field will monotonically decrease. Therefore, the contribution from the nuclear spins should cause a monotonic increase in the effective magnetic field, and therefore a monotonic increase in the spin's Larmor frequency. Modeling of quantum dots occupying a $^{28}$Si quantum well isotopically purified of nuclear spins to 800 ppm suggests that perhaps tens of $^{29}$Si nuclei may have a sizeable hyperfine coupling, for example such that $A_j/2\pi > \SI{10}{\kilo\hertz}$, to each electron spin \cite{Zhao_2019}. We may expect only a few of these to flip due to the change in the thermal equilibrium from \SI{20}{\milli\kelvin} to \SI{200}{\milli\kelvin}, and the resulting shift is an order-of-magnitude too small to account for the observed temperature dependence. Moreover, such nuclear spin flips would appear as discrete jumps in frequency. In contrast, we repeatedly observed the trend to be smooth.

Similar arguments hold for unpaired electron spins in the environment which may couple to the qubit spin through magnetic dipolar interactions. The gyromagnetic ratio of these spins would also be negative but with a magnitude that is three orders of magnitude larger. While essentially none of these environmental spins occupy the excited state at base temperature, a few percent of an ensemble may exist in the excited state at \SI{200}{\milli\kelvin}. Whether this change in magnetic environment meaningfully influences the qubits depends on the density and distance of unpaired spins which we currently cannot estimate.

The $g$-factor of an electron spin in a silicon quantum dot is renormalized by many factors including interface-induced spin-orbit coupling, micromagnet-induced spin-orbit coupling, the shape of the confinement potential, and crystal composition and strain. The renormalization from the micromagnet-induced spin-orbit coupling is on the order of \SI{10}{\kilo\hertz} and therefore any micromagnet dynamics are too small to account for the frequency shift via the $g$-factor. A $g$-factor change arising from a change in confinement potential can be seen as a general Stark shift effect, which we discuss below. It is possible that some degree of thermal expansion in the device, affecting both strain and spin-orbit coupling, may play a role in explaining the temperature-dependence of the Larmor frequency, although a rigorous estimate is extremely challenging to carry out given the scope of the microscopic details that must be accounted for. We note that \cite{Takeda_2018_freqshift} put forward a similar hypothetical origin by estimating the change in effective mass of the confined electron as a function of temperature \cite{Richard_2003_effmass}. This estimate relies on the validity of $\mathbf{k}\cdot\mathbf{p}$ model parameters at sub-Kelvin temperature scales.

The $g$-factor difference between valley-orbit states is worth consideration, since increased thermal energy may cause the excited valley state to become populated with greater probability during initialization \cite{Kawakami_2014, Ferdous_2018_valley}. However, we see no additional evidence, such as a second resonance peak, that such a valley state is being populated. Rather, the single resonance smoothly moves as temperature is increased.

It is well-known that electric fields may induce a Stark shift due to the spatial dependence of both the magnetic field $B_\textrm{tot,i}$ and $g$-factor $g_i$. A temperature-dependent electric field could thus indirectly cause a frequency shift. In our system, we could expect such a shift to be dominated by the micromagnet gradient particularly for an in-plane electric field that laterally displaces the electron. By this logic, it is reasonable that the frequency shifts we observe are generally larger than in other devices \cite{freer2017single,Savytskyy_2023_flipflop,zwerver2022qubits} where intrinsic spin-orbit coupling dominates. However, an electric field oriented perpendicular to the quantum well may also result in a frequency shift, depending on its magnitude. Where may the electric field inducing the Stark shift originate? Transport experiments through various quantum wells (similar to experiments in Si-MOS \cite{Tracy_2009_percolationtransition}) record an increased percolation density as a function of temperature which is believed to correspond to thermally activated defects causing electrostatic ``roughness'' (see Appendix~\ref{sec:Stark}). We observe the Larmor frequency shift over similar temperature scales, such that a thermally-activated electric field caused by rearranging environmental fluctuators may be a contributing factor.


We have not, however, been able to detect such a temperature-dependent electric field directly with the on-chip charge sensors. That is, Coulomb peak position and charge transitions within the quantum dot array move much less than \SI{1}{\milli\volt} with respect to their associated plunger gates as the temperature is raised from \SI{20}{\milli\kelvin} to \SI{250}{\milli\kelvin}. For a rough comparison, we may expect electric fields on the order of \SI{10}{\kilo\volt/\meter}, or equivalently plunger gate voltages of several millivolts, to lead to \SI{1}{\mega\hertz} shifts in the qubit frequency. We also find that temperature changes do not affect where Pauli spin blockade occurs in gate voltage space, a window of about \SI{2}{\milli\volt} along the interdot detuning axis. An electric field that simultaneously Stark shifts qubits while preserving these other charge-sensitive features is peculiar. We extend this discussion in Appendix~\ref{sec:Stark}. We also note that we do not observe any hysteresis in the temperature-dependence of the Larmor frequencies (see Fig.~\ref{fig:B2}).

In summary, temperature-dependence in the micromagnet may contribute to our observations, but observations of the heating effect in devices that do not contain a micromagnet suggest its role is not fundamental. A change in the magnetic environment of each spin due to residual hyperfine interaction is expected to be discrete and monotonic, unlike our observations, and likely smaller than the frequency shifts we report. Thermally-induced strain may also play a role, but it is very difficult to make an accurate numeric estimate. A temperature-dependent electric field which Stark-shifts the Larmor frequencies is compelling, given the thermal sensitivity of charge defects in the heterostructure. Our experiments do not reveal ``smoking gun'' evidence of such an electric field, but it is possible that a field manifests in a way such that our charge sensing apparatus is insensitive to it.

Lastly, we remark that it is difficult to conceive of a single mechanism that yields a nonmonotonous frequency shift. This perhaps points to an interplay between more than one relevant effect. Data from additional devices, especially those without a micromagnet, may help elucidate the dominant physics at play.

\section{Conclusion}

In summary, we have documented for the first time an explicit temperature-dependence in silicon spin qubit Larmor frequencies. The magnitude of the observed frequency shift is consistent with the heating effect that has long been associated with microwave control of individual spin states but may also be induced by baseband control signals. We find that this effect becomes substantially smaller as the temperature of the device is modestly increased, leading to reduced crosstalk and simpler calibration routines for achieving high-fidelity single-qubit control at \SI{200}{\milli\kelvin} without compromising coherence times. This makes microwave control of spins a viable means to continue scaling spin-based quantum processors in the near-term. While the microscopic origin of the temperature dependence remains elusive, we have highlighted some mechanisms that are consistent with our experiments. 


\begin{acknowledgments}
We acknowledge useful discussions with David DiVincenzo, Arne Laucht, Tom Watson, Maximilian Rimbach-Russ, Slava Dobrovitski, Ioan Pop, Pieter Eendebak, Nicky Wanningen, and members of the Vandersypen, Veldhorst, and Scappucci groups. This publication is part of the 'Quantum Inspire – the Dutch Quantum Computer in the Cloud' project (with project number [NWA.1292.19.194]) of the NWA research program 'Research on Routes by Consortia (ORC)', which is funded by the Netherlands Organization for Scientific Research (NWO). We acknowledge financial support from Intel Corporation. This research was sponsored by the Army Research Office (ARO) under grant numbers W911NF-17-1-0274 and W911NF-12-1-0607. The views and conclusions contained in this document are those of the authors and should not be interpreted as representing the official policies, either expressed or implied, of the ARO or the US Government. The US Government is authorized to reproduce and distribute reprints for government purposes notwithstanding any copyright notation herein. Data and analysis scripts supporting this work are available in the Zenodo repository, https://doi.org/10.5281/zenodo.7673892.

\end{acknowledgments}

\appendix
\section{Temperature dependence of rf reflectometry measurements}
\label{sec:readout}

\begin{figure*}
    \centering
    \includegraphics[width=\textwidth]{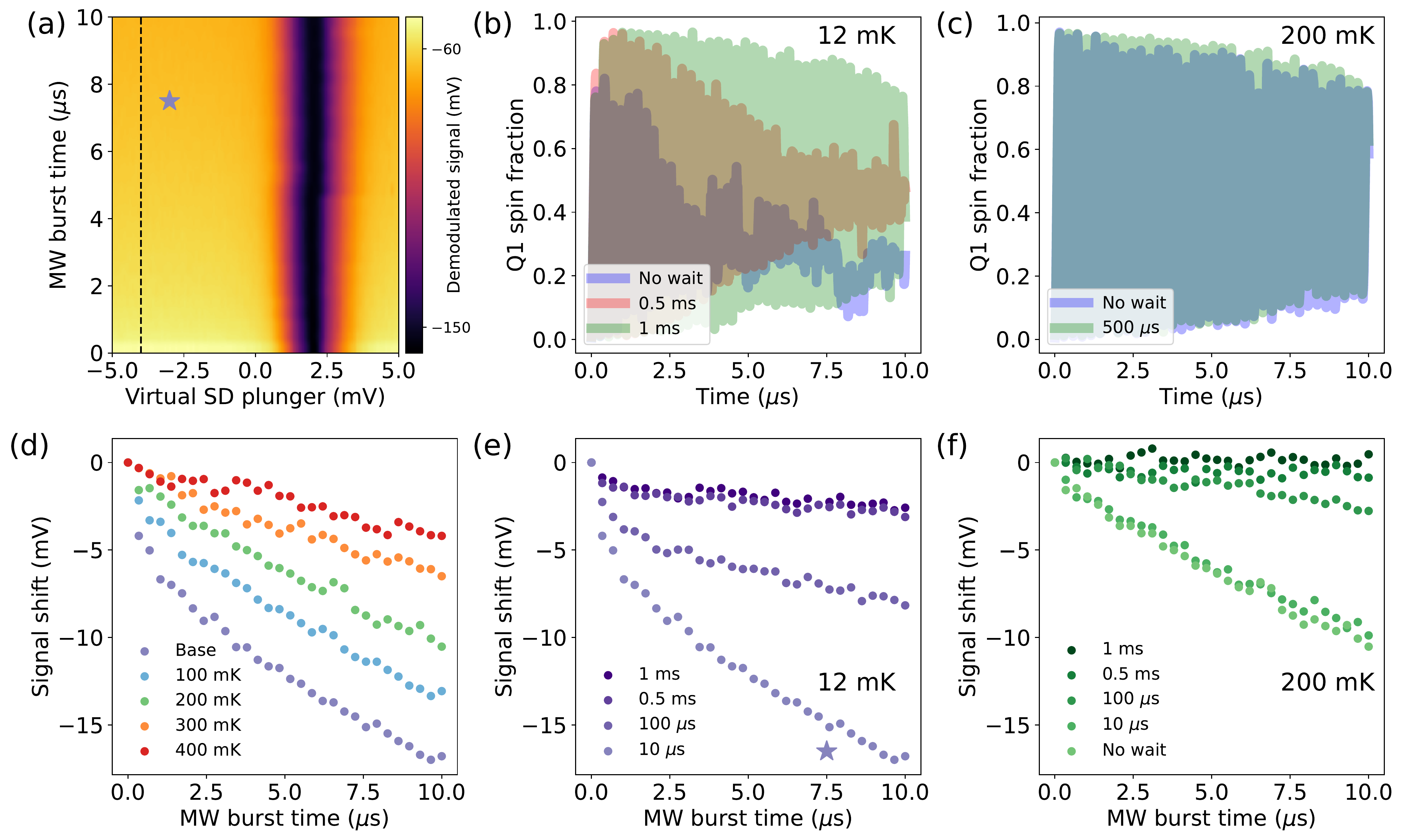}
    \caption{(a) A Coulomb conductance peak and its adjacent blockaded region measured using rf reflectometry at \SI{12}{\milli\kelvin} as a function of the virtualized sensing dot (SD) plunger voltage and a microwave burst of swept duration. The burst amplitude corresponds to $f_\textrm{Rabi}\approx\SI{1.3}{\mega\hertz}$. Due to the demodulation parameters used, the scalar signal at the conductance peak is invariant to microwaves up to $\pm\SI{2}{\milli\volt}$. (b-c) Rabi oscillations with $f_\textrm{Rabi}\approx\SI{3}{\mega\hertz}$ plotted to emphasize the apparent decay envelopes at \SI{12}{\milli\kelvin} and \SI{200}{\milli\kelvin}. The wait time indicates the idling duration between applying the resonant microwave tone and pulsing to the Pauli spin blockade region. (d) The integrated background signal along the line cut indicated in (a) taken at different mixing chamber temperatures. No wait time between microwave pulse and measurement is used, except at base temperature where a \SI{10}{\micro\second} pulse is employed. (e-f) The integrated background signal along the same line cut at \SI{12}{\milli\kelvin} and \SI{200}{\milli\kelvin} while different wait times between microwave pulse and readout were used.}
    \label{fig:A1}
\end{figure*}

While performing the experiments detailed in the main text, we observed that the rf reflectometry apparatus used to measure qubit spin states in this particular device also exhibits temperature dependence. While this dependence is not related to the qubit heating effects of primary focus in the main text, separating these two changes is not immediately trivial. Furthermore, as rf reflectometry has become a standard laboratory practice for measuring semiconductor spin qubits and quantum dot charge states more generally, we believe that careful consideration of its dependence on temperature will become an essential engineering consideration in more advanced devices.

The implementation of rf reflectometry in our device closely follows the ``Ohmic Approach'' documented in \cite{Liu_2021_rf}. We find that both the inductance $L$, which is implemented as an off-chip NbTiN high-kinetic inductance inductor, and the lead resistance $R_\textrm{lead}$, which includes the bond wire to the sample, the Ohmic contact, and the electron reservoir coupled to the sensing dot in series, are temperature-dependent. The reason for the inductor's temperature dependence was not studied in detail, but it was found to be a general property of inductors from the particular fabrication batch. Subsequent devices do not replicate this behaviour, so here we focus only on the consequences of the temperature dependence and not its origin. As discussed in the main text, some degree of temperature dependence in the lead resistance may be expected as a result of thermally-activated electrostatic disorder in the quantum well, but the series resistance through the Ohmic contact may also be a factor.

One of the two charge sensors is measured by demodulating a \SI{10}{}-\SI{20}{\micro\second} pulse of frequency $f_\textrm{res}$ and digitizing the resulting in-phase (I) and quadrature (Q) components of the signal. These two channels are converted into a single scalar value by rotating the IQ plane by an angle $\theta_\textrm{mod}$ such that the contrast between the two charge states of interest is maximized. A signal threshold $V_\textrm{th}$ is then set to enable single-shot spin readout. 

Temperature dependence in the rf reflectometry circuit is important because the probe frequency $f_\textrm{res}$, demodulating phase $\theta_\textrm{mod}$ and threshold $V_\textrm{th}$ are constant during a predefined experiment (e.g. measuring Rabi oscillations). However, a temperature dependence in $L$ and $R_\textrm{lead}$ would affect the optimal $f_\textrm{res}$, $\theta_\textrm{mod}$, and $V_\textrm{th}$ as microwaves and other control signals dissipate in the device. At best, this results in poorer readout fidelity and at worst can completely eliminate all readout visibility.

Fig.~\ref{fig:A1}(a) illustrates the consequences of microwave-induced heating at base (approx. \SI{12}{\milli\kelvin}) mixing chamber temperature on a Coulomb peak measurement. A long wait time of \SI{500}{\micro\second} is implemented between shots to maintain independence between measurements. Figs.~\ref{fig:A1}(d-f) show how the scalar background signal varies after a combination of microwave pulsing and waiting prior to signal integration at a variety of mixing chamber temperatures. The particular demodulation settings used during this set of experiments meant that the conductance peak signal was robust to microwave prepulsing within $\pm\SI{2}{\milli\volt}$ while the background signal was sensitive to changes in $\theta_\textrm{mod}$. In this case, the change in background signal directly corresponds to a lower charge state distinguishability.

Similar to the microwave-induced heating effect, we observe that the rf reflectometry signal is more robust as the device temperature is increased. However, we do not find that this effect becomes negligible, meaning that readout heating must still be controlled even at warmer device operation in order to maintain spin state visibility. The recovery of the original background signal amplitude as a function of wait time also illustrates how the cooling timescale becomes shorter when the mixing chamber temperature is increased. Although the signal at \SI{200}{\milli\kelvin} is modified by the microwave pulse, only \SI{100}{\micro\second} is required for the signal to recover. However, over \SI{0.5}{\milli\second} of waiting is necessary for the signal to recover at \SI{12}{\milli\kelvin}.

Fig.~\ref{fig:A1}(b) illustrate the effect of rf reflectometry heating on spin experiments. During a \SI{10}{\micro\second} single qubit Rabi oscillation experiment, some visibility is lost as the constant readout threshold no longer optimally distinguishes between charge states. In \cite{Philips_2022_6qubit}, a wait time was placed between qubit operation and measurement in anticipation of this readout heating effect. Otherwise, the loss of spin visibility gives the illusion of $T_2^\textrm{Rabi}$ decay. However, as the results in the main text illustrate, we find that this loss of visibility is also caused by the qubit frequencies shifting due to heat accumulated between shots.

Once again, we find that the most pragmatic solution is to work at an elevated temperature. At \SI{200}{\milli\kelvin}, the RLC resonator peaks are not majorly degraded but the qubit frequencies are much more stable throughout experiments. As such, maintaining readout visibility throughout experiments containing repeated microwave pulses is possible with no addition of intra-sequence wait-times whatsoever (see Fig.~\ref{fig:A1}(c)). The calibrated values of $f_\textrm{res}$, $\theta_\textrm{mod}$ and $V_\textrm{th}$ are more robust to signal dissipation during experiments as well. Ultimately, heating in the rf reflectometry circuit becomes the limiting experimental factor. During microwave-intensive experiments, such as single-qubit randomized benchmarking, where driving tones totaling tens of microseconds may be applied, a wait time between shots must be used or else the readout threshold $V_\textrm{th}$ will be so misaligned as to not provide any meaningful information due to steady heat accumulation.

We anticipate that a) using surface-mount inductors which exhibit less temperature dependence, and b) routing the rf reflectometry signal through a capacitively coupled gate as opposed to the galvanically connected ohmic contacts (see the ``Split-gate approach'' of \cite{Liu_2021_rf}) will lead to better thermal insulation between the RLC circuit and the quantum well. Ultimately, eliminating wait-times for ``cooling'' altogether will be essential for interleaving qubit measurement and control in future quantum error correction implementations.

\section{Probing charge defects and temperature-dependent Stark shifts}
\label{sec:Stark}

\begin{figure*}
    \centering
    \includegraphics[width=\textwidth]{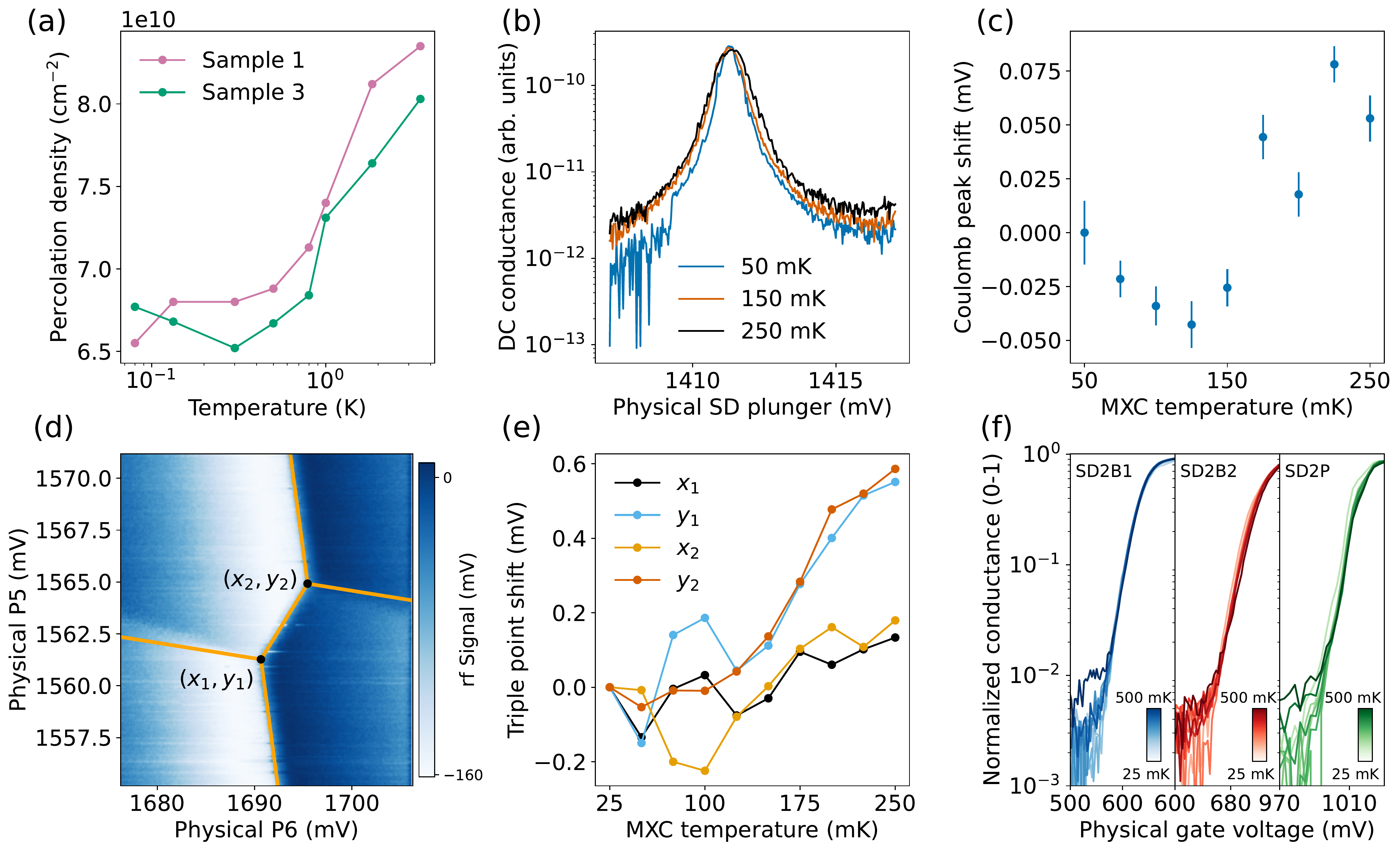}
    \caption{(a) Percolation density measurements as a function of temperature carried out on two $^{28}$Si/SiGe heterostructures nominally equivalent to that used for the qubit device studied here. (b) A selection of Coulomb peaks formed in SD2 measured in DC as a function of the physical sensing dot plunger voltage. These measurements were qualitatively unchanged when rf reflectometry was used instead. (c) Shifts in the Coulomb peak position in plunger gate voltage space ($V_\mathrm{G}$) when fitted to $\mathrm{d}I/\mathrm{d}V = A\cosh^{-2}[B(V_\mathrm{G}-V_0)]+C$ where $V_0$ is the location of the peak at \SI{50}{\milli\kelvin}, and $A$, $B$, and $C$ are other fitting constants. (d) A sample charge stability diagram of the $(1,3)-(0,4)$ interdot transition between quantum dots 5 and 6 taken at \SI{100}{\milli\kelvin}. An algorithm is used to extract the triple point coordinates plotted in (e). (e) The location of the two triple points labeled in (d) as a function of mixing chamber temperature. Charge stability diagrams were measured using rf reflectometry. (f) Comparison of threshold voltages in SD2 as a function of mixing chamber temperature, when using both sensing dot barriers (SD2B1, SD2B2) and the plunger (SD2P) to pinch off the accumulated channel. Measurements were taken using rf reflectometry to maximize sensitivity to any changes near the threshold.}
    \label{fig:B1}
\end{figure*}

Charge defects, or two-level fluctuators (TLFs), are ubiquitous in many solid state systems and are also expected to occupy the heterostructure in which the spin qubits studied here reside, as well as the dielectrics in the gate stack fabricated on top of the heterostructure. Although heavily researched for decades \cite{Paladino_2014}, much about their microscopic details remains unclear. It is generally well accepted that an ensemble of TLFs, each with a Lorentzian power spectral density, will cumulatively generate a $S(\omega)\propto 1/f^\alpha$ power spectral density with $\alpha\approx 1$ that is commonly cited as the origin of measured charge noise spectra \cite{dutta_1981_chargenoise}. Furthermore, the microscopic details of the TLFs may result in a variety of temperature dependencies at sub-Kelvin conditions \cite{Phillips_1987_TLS,Gustafsson_2013_thermal,Beaudoin_2015_chargenoise}.


Hall bar measurements shown in Fig.~\ref{fig:B1}(a) were carried out on the same $^{28}$Si/SiGe heterostructure as the six qubit device studied here and demonstrate that the percolation density increases as the device temperature is raised, particularly in the few-hundred millikelvin regime. This is believed to be the result of thermally-sensitive charge defects increasing the electrostatic disorder within the quantum well \cite{Tracy_2009_percolationtransition}. If this thermal activation of TLFs is correlated with the temperature-dependent Larmor frequencies we measure, we believe it is reasonable that an electric field-spin coupling facilitates this dependence.

It is well known that electric fields are able to shift electron spin Larmor frequencies in silicon quantum dots. Though often generically referred to as a Stark shift, the coupling may take place due to a spatially varying magnetic field, as in the case of an on-chip micromagnet, and a spatially varying $g$-factor dominated by interface-induced spin-orbit coupling. As these frequency shifts can be on the \SI{}{\mega\hertz} scale, it is reasonable to question whether a similar coupling may facilitate a temperature-dependent Stark shift.

Here we highlight three measurements that aim to uncover any correlation between the mixing chamber temperature and a change in electrostatic environment that may mediate the frequency shift and, consequently, motivate a microscopic origin of the ``heating effect''. First, we measure a Coulomb peak of Sensing Dot 2 (SD2) as a function of mixing chamber temperature by using a DC current. Figs.~\ref{fig:B1}(b-c) respectively show the Coulomb peaks and a fit of the peak locations along the sensing dot plunger axis versus temperature. During qubit readout, the peak location will move by about \SI{1}{\milli\volt} along this axis in the presence of a single electron transition between two dots in a radius of about \SI{200}{\nano\meter} around the sensing dot. The fact that we see no systematic peak movement of more than \SI{0.1}{\milli\volt} in the temperature range where the qubit frequencies shift by \SI{1}{\mega\hertz} implies there is little environmental charge rearrangement occurring in the vicinity of the sensing dot. This is consistent with Fig.~\ref{fig:A1}(a), measured via rf reflectometry, which also shows no Coulomb peak movement during microwave-induced heating.

Second, we measure the interdot charge transition between quantom dots 5 and 6 as a function of mixing chamber temperature. Figs.~\ref{fig:B1}(d-e) illustrate the fitted triple points and their movement in physical gate space as a function of temperature. A total movement of much less than \SI{1}{\milli\volt} is observed over the \SI{250}{\milli\kelvin} temperature range studied. This change is less than the width of the charge transitions themselves, and we remark that it is necessary to modify the plunger voltages by at least several millivolts in order to observe \SI{1}{\mega\hertz} Stark shifts in the qubit Larmor frequencies at a constant temperature. While this gives some evidence that the electrostatic environment near the quantum dots changes as a function of temperature, it does not form a convincing connection to the temperature-dependent frequency shift.

Third, we probe whether we can detect a change in the electrostatic roughness of the 2DEG reservoirs as a function of temperature which may provide a more global indication of the presence of TLFs near the quantum well. To do this, we use rf reflectometry to measure the conduction turn-on through SD2 as a function of all three gate voltages along the channel (2 barrier gates and 1 plunger gate). Temperature dependence in this turn-on characteristic may indicate that charge defects in the oxide between the gate and the $^{28}$Si/SiGe heterostructure activate and contribute to the net electric field at the quantum well. The rf measurement should be more sensitive than a dc current measurement since it may show a finite conductance before the 2DEG is fully accumulated. As shown in Fig.~\ref{fig:B1}(f), very little change in the threshold voltage is observed as a function of temperature, and no clear correlation emerged between this behaviour and the Larmor frequency shifts.

\begin{figure*}
    \centering
    \includegraphics[width=\textwidth]{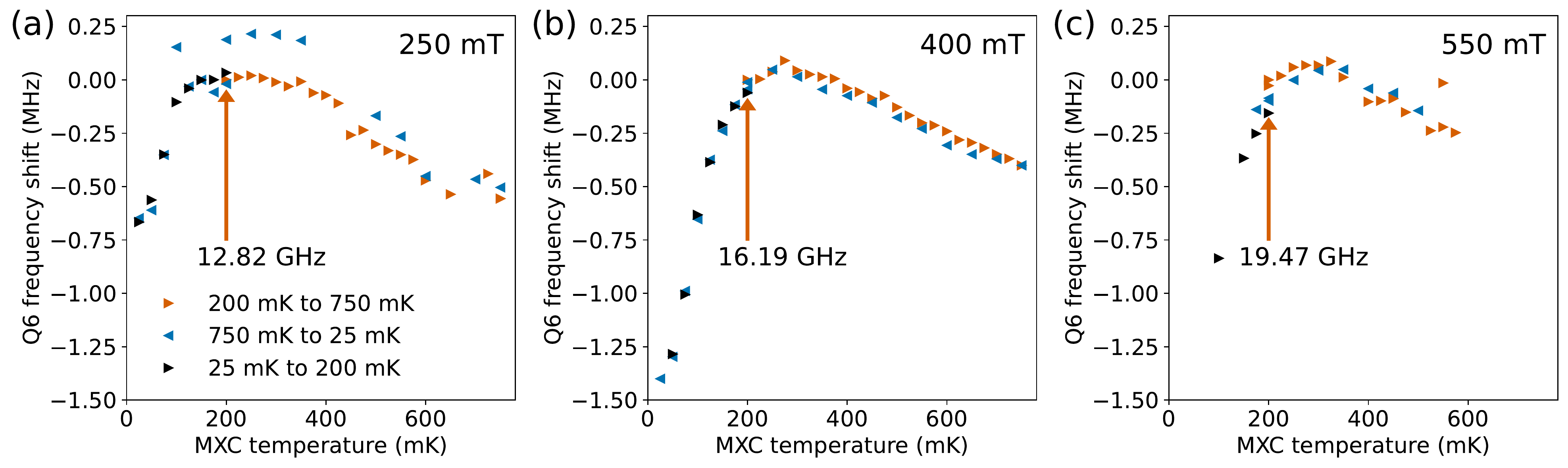}
    \caption{Tests for hysteresis in the temperature-dependent frequency shift of Qubit 6 carried out at external magnetic field settings of (a) \SI{250}{\milli\tesla}, (b) \SI{400}{\milli\tesla}, and (c) \SI{550}{\milli\tesla}. In each case, the absolute qubit resonance frequency is indicated at \SI{200}{\milli\kelvin} which also serves as a reference for the magnitude of the frequency shift. The device is positioned such that it experiences approximately 80\% of the nominal external field quoted. Accordingly, the Larmor frequency scales linearly with the external field at a rate of \SI{22.2}{\giga\hertz/\tesla}, indicating that the micromagnet stray field is approximately constant for all experiments. To test for hysteresis, the temperature is swept from \SI{200}{\milli\kelvin} to \SI{750}{\milli\kelvin} in \SI{25}{\milli\kelvin} steps, then down to \SI{200}{\milli\kelvin} in \SI{50}{\milli\kelvin} steps, then down to \SI{25}{\milli\kelvin} in \SI{25}{\milli\kelvin} steps, and finally back up to \SI{200}{\milli\kelvin} in \SI{25}{\milli\kelvin} steps. In (a) and (c), a low-frequency electrical TLF coupled to both the qubit and adjacent charge sensor. Experiments where readout quality is compromised entirely are omitted for clarity.}
    \label{fig:B2}
\end{figure*}

A final series of tests were carried out to detect whether the qubit frequency shifts exhibited a hysteresis effect with respect to temperature. As microscopic models of TLFs commonly include asymmetry in the constituent bi-stable states, hysteresis of the frequency shift with respect to temperature could provide an indication of this structure. Repeated experiments at three different external field settings, of which a few are included in Fig.~\ref{fig:B2} yield no systematic hysteresis effect. Although jumps in the qubit frequency and Coulomb peak tuning occur occasionally, the underlying temperature dependence remains robust, and repeated experiments generally do not reproduce the frequency shifts originating from bi-stable behaviour. Simultaneous jumps of both the Coulomb peak and qubit frequencies are likely the result of nearby charge jumps in the system, whereas jumps in qubit frequency without an associated Coulomb peak shift may arise due to nuclear spin flips \cite{Zhao_2019}. The characteristic temperature dependence of the qubit frequencies is observed regardless of whether these discrete jumps occur during an experiment. We remark that the magnitude of the temperature-dependent frequency shift is relatively larger at higher magnetic fields which is consistent with the hypothesis that the temperature-dependence originates with the electron $g$-factor. However, potential changes in the micromagnet magnetization do not let us make conclusive statements from this data. The reported microwave-induced frequency shifts in devices without a micromagnet provide the most compelling evidence that the $g$-factor itself is the fundamentally temperature-dependent quantity \cite{freer2017single,zwerver2022qubits,Savytskyy_2023_flipflop}.

In summary, our experiments do not rule out an electrostatic or TLF contribution to the Larmor frequency temperature dependence. However, the exploratory results presented here certainly do not uncover a ``smoking gun'' of a significant effect. Ongoing research into the concentration and microscopic details of the dominant charge defects in semiconductor quantum wells will help contextualize their possible role in the ``heating effect''.

\section{Electron temperature measurements}
\label{sec:Elec_temp}

We measure estimates for the electron temperature on both sides of the device by measuring Coulomb diamonds through the two sensing dots at base mixing chamber temperature. These yield estimates of $\SI{180}{}\pm\SI{10}{\milli\kelvin}$ and $\SI{200}{}\pm\SI{10}{\milli\kelvin}$ for the 2DEG reservoirs surrounding sensing dots 1 and 2 respectively. Estimates of electron temperature made via reservoir-dot charge transitions yield similar, but somewhat higher, estimates of \SI{232}{\milli\kelvin} and \SI{214}{\milli\kelvin} for sensing dots 1 and 2 respectively.

Although these temperatures are similar to the ``sweet spot'' appearing in Fig.~\ref{fig:2}, it is unclear whether this is a meaningful or coincidental correlation. However, it does seem reasonable that raising the mixing chamber temperature to at least the electron temperature facilitates faster device thermalization as various dissipative control signals are sent to the device, because electron diffusion may facilitate thermalization of the quantum dot environment when the phononic and electronic environments share a similar temperature. This may, in part, explain the observations in Fig.~\ref{fig:7} where microwave induced heating is greatly reduced at temperatures similar to (or above) the electron temperature floor.

\section{Spin coherence at 200 mK}
\label{sec:coherence}

\begin{figure}
    \centering
    \includegraphics[width=\linewidth]{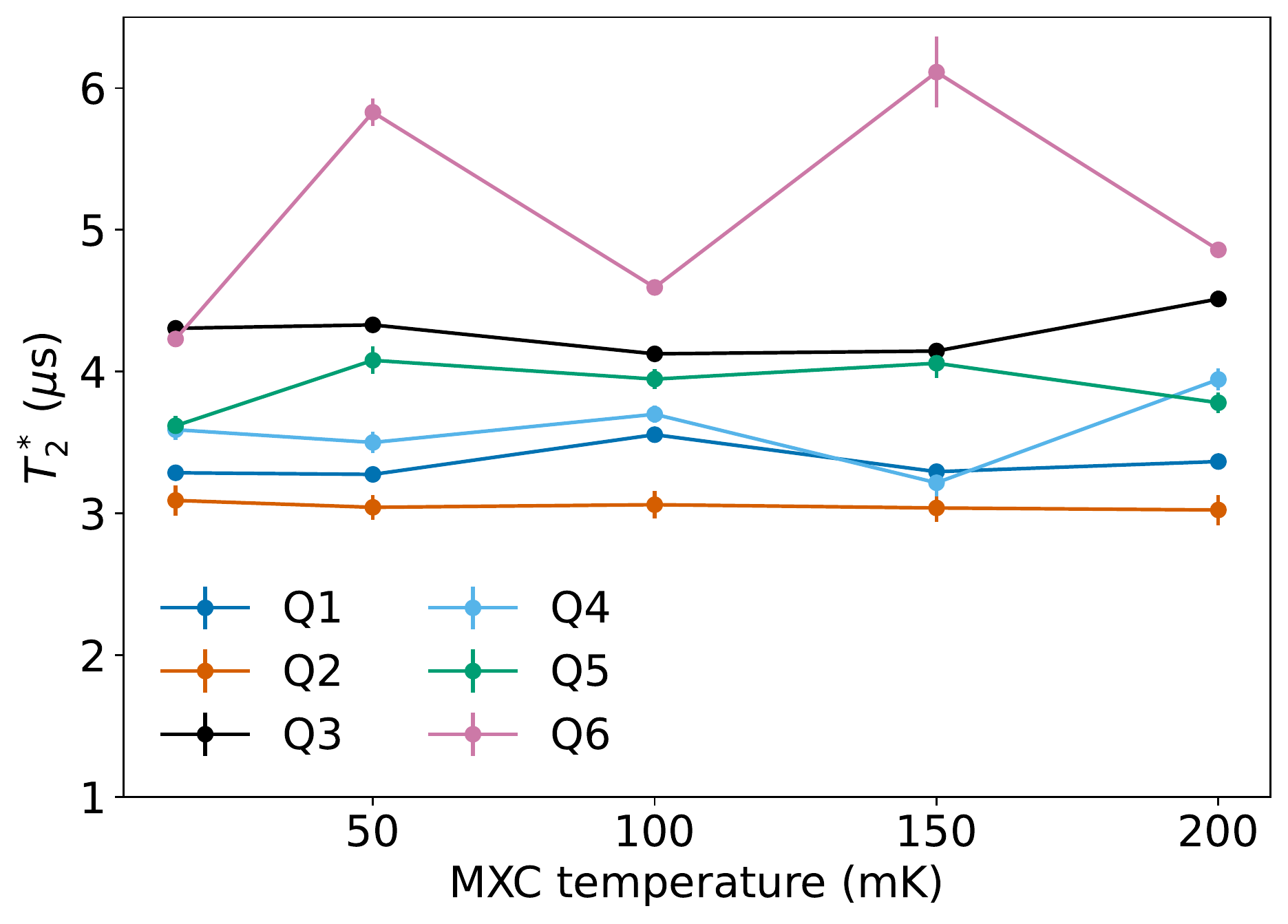}
    \caption{$T_2^*$ measurements extracted from the same Ramsey experiments used to measure Larmor frequency temperature dependence in Fig.~\ref{fig:2}. Each single-qubit Ramsey measurement is averaged for approximately 8 minutes.}
    \label{fig:D1}
\end{figure}

Fig.~\ref{fig:D1} illustrates how increasing the mixing chamber temperature from base to \SI{200}{\milli\kelvin} has virtually no impact on the $T_2^*$ values for all six qubits. The values reported here are directly comparable to previous measurements on the same device when no prepulsing was used prior to qubit control \cite{Philips_2022_6qubit}. In contrast, when a \SI{4}{\micro\second} microwave prepulse of normal gate-operation amplitude was applied prior to the measurement, $T_2^*$ values fell to \SI{1}{}-\SI{3}{\micro\second} across the array. From this, we conclude that directly raising the mixing chamber temperature to overcome the heating effect is superior to prepulsing.

A full analysis of how qubit coherence times, noise spectra, and spin readout vary over a wider range of temperature will be included in a separate publication. In the context of this work, we highlight that the modest increase in device temperature to \SI{200}{\milli\kelvin} used to overcome contextual operation does not come at the cost of compromising coherence times or single-qubit gate fidelities.

\end{document}